\documentclass{article}
\usepackage{arxiv}

\usepackage[utf8]{inputenc} % allow utf-8 input
\usepackage[T1]{fontenc}    % use 8-bit T1 fonts
\usepackage{hyperref}       % hyperlinks
\usepackage{url}            % simple URL typesetting
\usepackage{booktabs}       % professional-quality tables
\usepackage{amsfonts}       % blackboard math symbols
\usepackage{nicefrac}       % compact symbols for 1/2, etc.
\usepackage{microtype}      % microtypography
\usepackage{lipsum}		% Can be removed after putting your text content
\usepackage{graphicx}
\usepackage{natbib}
\usepackage{doi}

\title{Perceptually Lossless Tactile Texture Synthesis with Compact Spectral Envelope Models}

\author{ \href{https://orcid.org/0000-0002-8990-9629}{\includegraphics[scale=0.06]{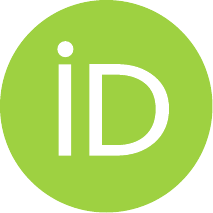}\hspace{1mm}Jagan K.~Balasubramanian}\\
	Department of Cognitive Robotics,\\
	Faculty of Mechanical Engineering\\
    Delft University of Technology\\
	Delft, 2628 CC \\
	\texttt{j.krishnasamybalasubramanian@tudelft.nl} \\
	\And
	\href{https://orcid.org/0000-0003-2156-1504}{\includegraphics[scale=0.06]{orcid.pdf}\hspace{1mm}Yasemin Vardar}\thanks{To whom correspondence should be addressed.}  \\
	Department of Cognitive Robotics,\\
	Faculty of Mechanical Engineering\\
    Delft University of Technology\\
	Delft, 2628 CC \\
	\texttt{y.vardar@tudelt.nl} \\
}

\renewcommand{\shorttitle}{\textit{arXiv} Template}

\hypersetup{
pdftitle={Perceptually Lossless Tactile Texture Synthesis with Compact Spectral Envelope Models},
pdfsubject={Haptics},
pdfauthor={Jagan K.~Balasubramanian, Yasemin~Vardar},
pdfkeywords={Texture representation and synthesis, Data-driven modeling, Haptics and Haptic Interfaces, Tactile Perception},
}

\begin{document}
\maketitle

\begin{abstract}
	Modern audio-visual media rely on compact representations for efficient storage and transmission, whereas realistic digital touch still depends on high-resolution tactile recordings. Existing approaches for representing tactile signals constrain manipulation and limit the generation of new content. Here, we introduce two compact representations—spectral beta and spectral slope—that capture the temporal spectral structure of finger-surface friction signals while preserving perceptually relevant information. Spectral beta models spectral skewness using a two-parameter beta distribution, whereas spectral slope approximates the spectrum with an asymmetric bandpass filter defined by low- and high-pass orders. We evaluated these representations in a perceptual study with 14 participants using five virtual textures rendered on a friction-modulation display and compared them with physical textures and high-fidelity reproductions of recorded signals. Spectral beta achieved perceptual similarity ratings comparable to those of the original high-fidelity reproductions. Regression analysis further showed that matching spectral energy across nine critical frequency bands was the strongest predictor of perceived realism. Together, these findings suggest that tactile texture perception depends primarily on fundamental temporal spectral patterns and that modeling these patterns is sufficient for perceptually realistic rendering. These results establish an efficient and scalable framework for haptic compression, communication, and synthetic texture generation.
\end{abstract}

% keywords can be removed

\section{Introduction}
%The \Firstpage command is used to format the first page text column size. The same size will be maintained for subsequent paragraphs until the \Endparasplit or \Parasplit command is encountered.
Today's digital technologies enable users to stream music, view high-resolution videos, and capture photographs with remarkable realism. In contrast, the digital processing of tactile information is largely limited to simple vibrations, falling short of the rich sensations experienced in the physical world. One reason for this gap is the decades of intensive research in audio-visual domains, which have produced perceptually lossless representation techniques for efficient storage, compression, and content generation~\cite{ji2020comprehensive, duong2025evaluation,tavares2025traditional,zhou2024survey}. %These advances now enable the synthesis of images, videos, and music from learned representations conditioned on user-provided text prompts~\cite{radford2021learning,agostinelli2023musiclm}. 
Extending similar capabilities to touch is increasingly important, as touch is a primary mode of interaction with both the physical and digital worlds. Achieving compact and perceptually lossless representations for tactile signals, and enabling their efficient synthesis and rendering, will be essential for bringing haptics to the level of audio-visual media and for realizing immersive multimodal experiences~\cite{steinbach2018haptic, noll2020rate}.

Conventional approaches to digital touch rely either on generating tactile signals from predetermined parameters~\cite{meyer2020interfaces,friesen2021building,burns2024single} or on directly recording and replaying raw signals from finger- or tool-surface interactions~\cite{okamura1998vibration, culbertson2014modeling, grigorii2021data}. The former approach can produce a wide range of tactile effects with minimal effort, but often struggles to reproduce specific, realistic sensations. In contrast, the latter can capture high realism, but requires specialized hardware and expertise, making it difficult and costly to scale across the diversity of real-world surfaces.  Moreover, transmitting raw tactile signals alongside audio and video streams imposes substantial bandwidth demands~\cite{steinbach2018haptic,zhang2026tactile}.
As such recordings are tied to specific textures and interaction conditions~\cite{fagiani2012contact,manfredi2014natural,culbertson2015should}, they are also difficult to manipulate, limiting the synthesis of new tactile experiences.

%During interactions with real surfaces, either through tool-tips~\cite{culbertson2014modeling} or bare fingers~\cite{balasubramanian2024sens3}, the generated vibration or friction signals are captured using specialized equipment under controlled exploration conditions~\cite{balasubramanian2024sens3} and rendered directly on haptic devices~\cite{culbertson2014modeling, fielder2019novel, vardar2025multimodal,cozcolluela2025generating}. However, acquiring and storing such data —which varies with tool-tip or finger normal force~\cite{fagiani2012contact} and scanning speed~\cite{callier2015kinematics} —is costly and labor-intensive~\cite{okamoto2012lossy,steinbach2018haptic}, making it impractical to capture and render the full range of real-world textures. 

\begin{figure}[!ht]
    \centering
    \includegraphics[width=\linewidth]{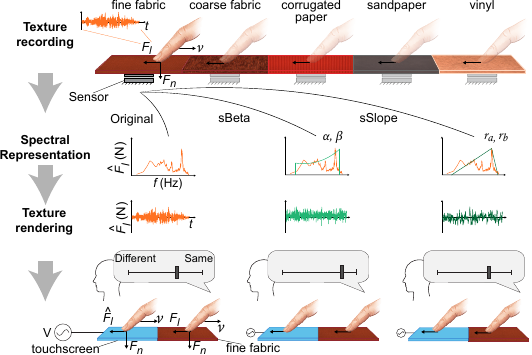}
    \caption{\textbf{Overview of the proposed texture representation and evaluation process.} Contact forces were recorded as a participant slid their finger on five surfaces. The recorded forces ($F_l$) were analyzed in the frequency domain and represented using spectral shape parameters ($\alpha$ and $\beta$) for the spectral beta (sBeta), and roll-off rates ($r_a$ and $r_b$) for the spectral slope (sSlope). The resulting signals ($\hat{F_l}$) were rendered via electrovibration. Participants subsequently rated the similarity between the rendered and real textures.}
    \label{fig: concept}
\end{figure}

Consequently, researchers have explored a wide range of data-driven representation methods for compressing and synthesizing tactile data, particularly for recorded vibrations or frictional signals associated with textures. Early approaches drew on audio coding techniques, including linear predictive coding~\cite{romano2011creating, shin2020hybrid}, autoregressive coefficients~\cite{culbertson2014modeling}, and Mel-frequency cepstral coefficients (MFCC)~\cite{hassan2020authoring}. %These methods compress raw interaction signals and enable synthesis through coefficient interpolation in perceptual or affective spaces linked to surface descriptors or exploratory behavior. 
Other approaches represent textures via spatially encoded friction maps, known as texels~\cite{meyer2016tactile}, or dominant spectral peaks~\cite{fielder2019novel}. %Extending these approaches, some studies model the temporal spectral patterns of haptic signals using Gaussian distributions ~\cite{friesen2021building,burns2024single}. 
%In parallel, image-based approaches employ generative adversarial networks (GANs)~\cite{ujitoko2018_gan,cai2022_gan} to infer contact friction or acceleration data from visual inputs.
Additional methods, such as perceptually bitrate-scalable coding and sparse linear prediction, have focused on transmission efficiency by compressing tactile signals while preserving perceptually relevant information~\cite{chaudhari2014perceptual,hassen2020pvc}.
 
Despite their ability to represent and reproduce recorded vibration or friction signals, existing approaches fall short of providing a perceptually meaningful basis for texture representation. Such representations could not only enable the manipulation of existing signals or the synthesis of new textures without additional recordings, but also provide insight into how humans perceive textures. Methods with many parameters, such as spatial friction-map techniques~\cite{meyer2016tactile}, can capture rich signal detail but make it difficult to establish generalizable relationships for texture perception and synthesis. Conversely, approaches with fewer parameters, such as temporal audio-coding~\cite{hassan2020authoring} and spectral peak models~\cite{fielder2019novel}, are more compact but largely texture specific, limiting their ability to generalize across the wide variety of real-world surfaces. This limitation underscores the need for a compact, perceptually grounded representation that supports both efficient synthesis and generalization to novel tactile textures.

To address these challenges, we propose two compact representations for perceptually lossless tactile texture synthesis: spectral beta \textbf{(sBeta)} and spectral slope \textbf{(sSlope)} (Figure~\ref{fig: concept}). These models capture the shape of the temporal spectrum of recorded finger-surface friction signals. %and generate virtual textures on a friction-modulation display that are perceptually similar to the original surfaces in terms of roughness and friction. 
sBeta represents the spectrum's profile using a beta distribution with shape parameters $\alpha$ and $\beta$, while sSlope models the spectrum using an asymmetric narrow-bandpass filter defined by its low- and high-pass filter orders. 

The validity of these representations was evaluated in a human-participant experiment, in which fourteen participants interacted with five virtual textures rendered on an electrostatic friction-modulation display~\cite{bau2010teslatouch} using both the  
proposed methods and existing approaches (auto regression~\cite{culbertson2014modeling}, mel frequency cepstral coefficients~\cite{hassan2020authoring}, and spectral peak~\cite{fielder2019novel}), and were rated their realism relative to the corresponding physical surfaces (fine fabric, coarse fabric, corrugated paper, sandpaper, and vinyl). 

Textures represented by sBeta were perceived as the most realistic for all the tested surfaces, even though the correlation between rendered and original friction signals was moderate. Linear regression further revealed that representations matching spectral energy across nine critical frequency bands predicted higher perceptual similarity. These findings demonstrate that simple, low-dimensional representations can preserve the perceptual essence of tactile textures. 

\section{Results}

\begin{figure*}[!ht]
    \centering
    \includegraphics[width=\linewidth]{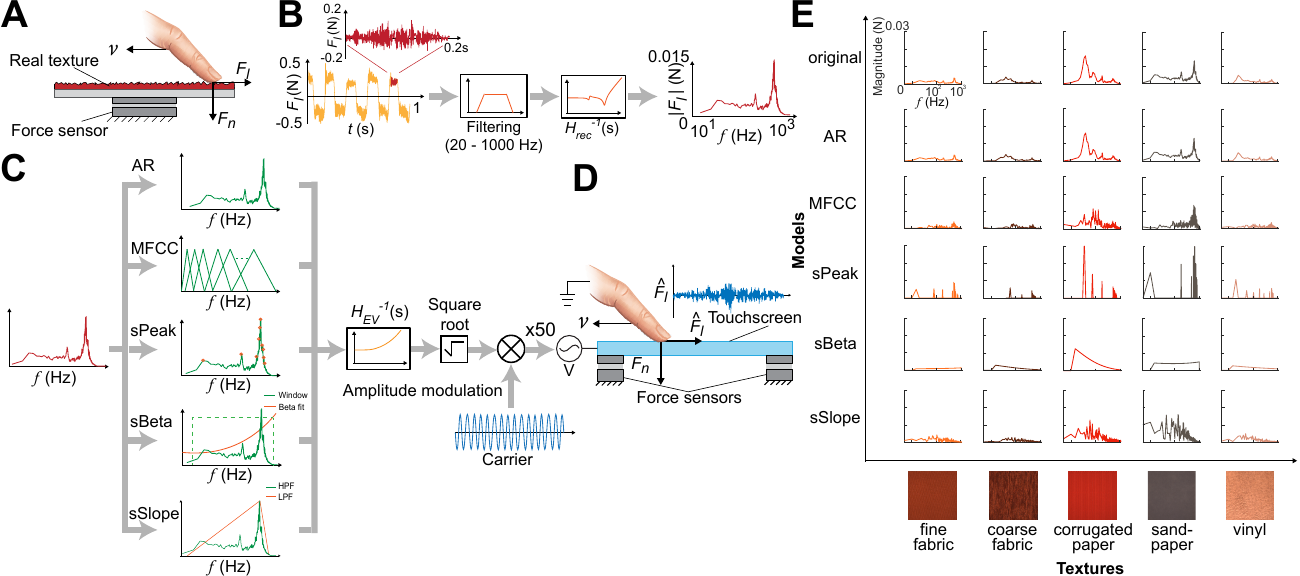}
    \caption{\textbf{Overview of the texture recording, representation, and rendering process.} (A) Contact forces were recorded as a participant slid their finger across surfaces mounted on an acrylic base attached to a force sensor. The interaction was controlled at a normal force ($F_N$) = 0.4~N and sliding speed ($\nu$) = 80~mm/s. (B) The recorded lateral force signals, $F_{L}$, were segmented during lateral sweeps (red), filtered within the human exploration bandwidth (10–1000~Hz), compensated for the setup's response ($H_{rec}^{-1}(s)$), and transformed to the frequency domain. (C) Texture representations were computed using autoregression (AR), Mel-frequency cepstral coefficients (MFCC), spectral peak (sPeak), and the proposed spectral beta (SBeta) and spectral slope (sSlope). Model outputs were compensated for the electrovibration response ($H_{EV}^{-1}(s)$) and amplitude-modulated with a high-frequency carrier (7000~Hz) following a square root operation. (D) The processed signals were rendered via electrovibration, exciting the touchscreen to generate lateral friction ($\hat{F_{L}}$). (E) Frequency-domain outputs of the different texture representation methods are compared with the original lateral force spectra from finger–surface interactions across the five tested textures.}
    \label{fig: modelOut}
\end{figure*}

Contact friction signals were recorded as a participant (first author) slid their fingertip across five different surfaces: fine fabric, coarse fabric, corrugated paper, sandpaper, and vinyl (Figure~\ref{fig: modelOut}A). The recorded signals were then compensated for finger motion and recording setup dynamics (Figure~\ref{fig: modelOut}B) and modeled using well-established representation methods---autoregression (AR), Mel frequency cepstral coefficients (MFCC), spectral peak (sPeak)---as well as our proposed compact models, spectral beta (sBeta) and spectral slope (sSlope) (Figure~\ref{fig: modelOut}C). Texture signals synthesized from these representations were subsequently rendered on an electrovibration display (Figure~\ref{fig: modelOut}D). During rendering, the resulting contact friction signals generated as the participant interacted with the display were recorded. Detailed procedures regarding texture recording, representation, and rendering are provided in Materials and Methods. 

\begin{figure}[!ht]
    \centering
    \includegraphics[width=0.9\linewidth]{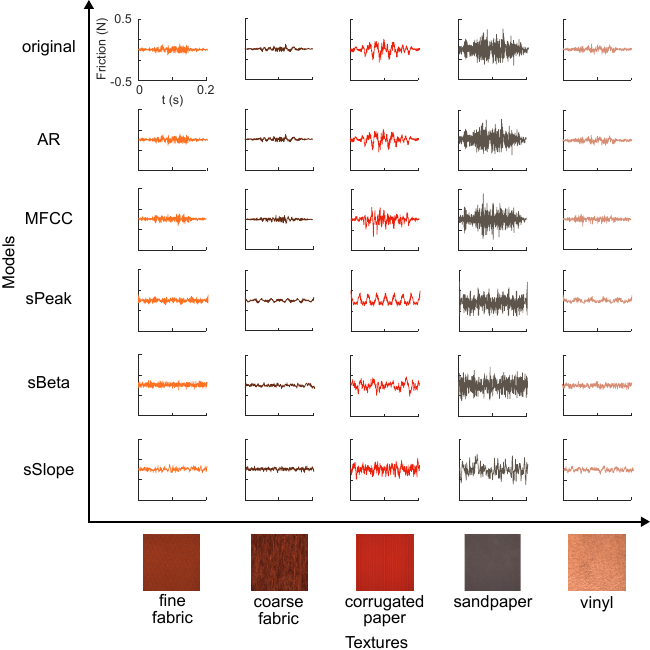}
    \caption{Time-domain outputs of the texture representation methods—auto regression (AR), Mels frequency cepstral coefficients (MFCC), spectral peak (sPeak), spectral beta (sBeta), and spectral slope (sSlope)—compared with the original lateral force spectra from finger–surface interactions for the five tested textures.}
    \label{fig: modelOut_time}
\end{figure}

Comparisons between the original and the rendered friction signals measured from the same participant are shown in Figure~\ref{fig: modelOut}E and Figure~\ref{fig: modelOut_time}. As demonstrated, the contact friction signals varied substantially across surfaces. Fine fabric, coarse fabric, and vinyl exhibited lower overall spectral energy than the corrugated paper and sandpaper. Moreover, coarse fabric, corrugated paper, and vinyl showed dominant spectral power at lower frequencies, whereas fine fabric and sandpaper displayed broader spectral distribution extending toward higher frequencies. The different representation methods preserved these spectral characteristics to varying degrees. AR produced spectra closely matching the original signals, whereas spectral-peak representations captured only limited spectral information. MFCC representation better preserved higher frequency components, while sBeta and sSlope approximated the overall shape of the frequency spectrum.

To evaluate the validity of our representations, we conducted a user study in which 14 naive participants interacted with the virtual textures and rated their realism relative to the corresponding physical surfaces in terms of perceived friction and roughness using a 7-point scale (0 = completely different, 6 = completely same). Each texture–representation condition was repeated five times. During the experiments, contact friction signals generated on the electrovibration display were recorded. Detailed experimental procedures are provided in the Materials and Methods section. 

\begin{figure}[!ht]
    \centering
    \includegraphics[width=\linewidth]{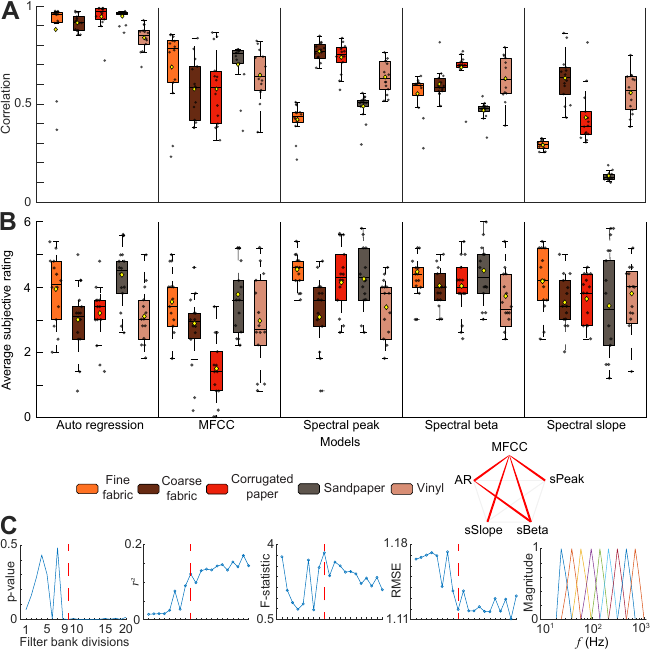}
    \caption{\textbf{Experimental results.} (A) Spectral correlation between rendered and original friction signals across 14 participants, five textures, and five representations. (B) Subjective similarity ratings, averaged over five repetitions, comparing rendered and real textures; filled circles denote individual participants, diamonds indicate mean across participants; the red lines next to the legends denote statistically significant differences between the models(p $<$ 0.05, Bonferroni-corrected). (C) Linear regression metrics (p-value, $r^2$, F-statistic, and root mean square error (RMSE)) relating standardized subjective ratings to the standardized spectral energy difference between rendered and original signals. Regression features were derived by progressively subdividing the spectral energy difference into multiple frequency bands. The dashed red line indicates that matching energy across nine critical frequency bands yields high perceptual similarity for the tested textures. The final panel illustrates these nine frequency bands.}
    \label{fig: result}
\end{figure}

% Finger–touchscreen contact forces and subjective similarity ratings (0 = different, 6 = same) were collected from 14 participants as they compared rendered textures with their real-world counterparts.  Additional details are provided in the supplementary material.
The spectral correlation between the original recordings and measured contact friction signals during the rendering is shown in Figure~\ref{fig: result}A. To quantify spectral fidelity, we computed correlations between the frequency-domain spectra of rendered signals and the corresponding original recordings, averaged across participants and repetitions. Frequency-domain analysis was used because variability in finger motion introduces phase shifts that make time-domain comparisons unreliable. 
The AR model preserved the highest spectral fidelity during electrovibration rendering, achieving correlations above 0.9 for all textures except vinyl (0.85). Other representations showed texture-dependent performance. MFCC representation achieved moderate correlations (0.6 – 0.7) and performed better for fine textures, such as fine fabric and sandpaper, whereas the sPeak method yielded the highest similarity (0.7 – 0.8) for coarse textures, including coarse fabric and corrugated paper. In contrast, sBeta produced lower correlations than sPeak, and sSlope showed the lowest correlations across textures, particularly for fine textures. 

Average subjective similarity ratings (Figure~\ref{fig: result}B) showed that sBeta yielded the most consistent perceptual similarity across textures, performing well for both fine and coarse surfaces. Fine textures, such as fine fabric and sandpaper, were generally rated highly (ratings $>$ 4) across most methods, with the exception of MFCC for all textures and sSlope for sandpaper. For coarse textures, including coarse fabric and corrugated paper, sBeta achieved the highest similarity ratings ($\approx$4), followed by sPeak and sSlope, whereas AR and MFCC scored lower. AR performed better on fine than on coarse textures; MFCC consistently produced the lowest ratings for coarse textures; sPeak produced ratings similar to sBeta, except for coarse fabric; and sSlope exhibited the greatest variability across textures. 

After confirming normality of the subjective similarity ratings using the Kolmogorov–Smirnov test, a repeated-measures ANOVA was conducted on the similarity ratings. The results revealed a statistically significant effect of representation type on subjective similarity ratings (F(4, 52) = 19.997, p $<$ 0.001). Post-hoc pairwise comparisons with Bonferroni correction showed that MFCC differed significantly from AR, sPeak, sBeta, and sSlope (p $<$ 0.001), whereas AR differed significantly from sBeta (p $<$ 0.05) (Figure~\ref{fig: result}B). Texture-specific comparisons (Figure~S5) revealed no significant differences in representations for sandpaper and vinyl. For fine fabric, a significant difference was observed between sPeak and MFCC (p $<$ 0.05). For coarse fabric, sBeta differed significantly from AR (p $<$ 0.05) and MFCC (p $<$ 0.01). For corrugated paper, MFCC differed significantly from AR (p $<$ 0.001), sPeak (p $<$ 0.001), sBeta (p $<$ 0.001), and sSlope (p $<$ 0.001), AR differed significantly from sPeak (p $<$ 0.05) and sBeta (p $<$ 0.05), and sPeak differed significantly from sSlope (p $<$ 0.05).

To identify which spectral regions contributed most strongly to perceived realism, we modeled subjective similarity ratings as a function of band-wise spectral energy differences between the original texture and the measured contact forces. Linear regression analysis (Figure~\ref{fig: result}C) revealed that preserving spectral energy across nine logarithmically spaced frequency bands was sufficient to predict subjective ratings, with model performance plateauing when additional bands were included. Prior to analysis, friction signals and ratings were averaged across repetitions, and energy differences between original and rendered textures were calculated. As participants adjusted the gain of the rendered outputs to match perceived friction with the real textures, the spectral energy differences and average ratings were standardized for each participant and representation. These results suggest that perceived texture realism depends strongly on preserving coarse spectral energy distributions across a limited number of frequency bands rather than reproducing the full spectral detail of the original signals (see Supporting Information for more
details).

\section{Discussion}
% Mention why AR has the highest correlation.
% Fine textures are rendered well with electrovibration. Also, add a citation. Mention that the overall response was dominated by the paper in the analysis.

This work introduces two compact texture representations, spectral beta (sBeta) and spectral slope (sSlope), that model the temporal spectral structure of tactile texture signals to enable efficient synthesis while preserving perceptually relevant information. In a perceptual study with 14 participants, these representations were evaluated against established methods- autoregression (AR), Mel frequency cepstral coefficients (MFCC), and spectral Peaks (sPeak)-by comparing virtual textures rendered on a friction-modulation display with their corresponding physical surfaces. Among these approaches, sBeta achieved perceptual similarity ratings comparable to those of the high-fidelity AR model (Figure~\ref{fig: result}B), despite requiring only two parameters. These findings suggest that preserving coarse spectral structure is sufficient to reproduce perceptually realistic textures and may provide a foundation for efficient haptic compression and synthesis.

The results showed that sBeta achieved perceptual similarity ratings comparable to, and in some cases exceeding, those of the other compact representations (Figure~\ref{fig: result}B). One possible explanation is that the recorded textures exhibited spectral skewness toward either narrow low-frequency bands or broader high-frequency regions (Figure~\ref{fig: modelOut}E). By modeling spectral structure on a normalized logarithmic frequency scale, sBeta captures this skewness efficiently with two parameters, providing finer resolution at low frequencies while more compactly representing broad high-frequency content. This advantage is likely reinforced by the characteristics of electrovibration perception, which is more sensitive to low frequencies with smaller just-noticeable differences~\cite{bau2010teslatouch}. Coarse textures tend to concentrate energy within narrow, low-frequency bands, whereas fine textures distribute energy across broader high-frequency ranges. By capturing both regimes within a compact representation, sBeta is able to reproduce tactile textures with high perceptual similarity ratings. 

\begin{figure}[!ht]
    \centering
    \includegraphics[width=\linewidth]{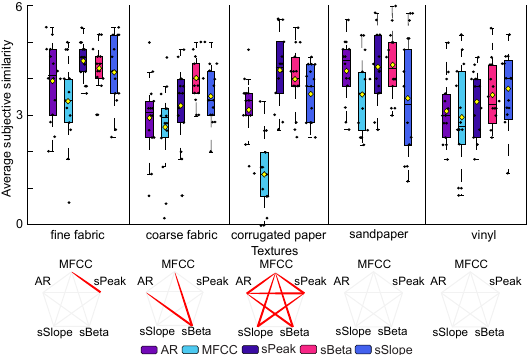}
    \caption{Subjective similarity ratings averaged over five repetitions. Filled circles represent each participant’s mean rating (n = 14), and the yellow cross shows the overall mean. Double asterisks (**) with brackets indicate statistically significant differences (p $<$ 0.05, Bonferroni-corrected).}
    \label{fig: Result_tex}
\end{figure}

Overall, perceptual similarity results showed that textures rendered with sPeak, sBeta, and sSlope were largely indistinguishable, whereas textures rendered with MFCCs differed substantially from the original textures (Figure~\ref{fig: result}B). Although AR achieved high spectral correlation with the original signals (Figure~\ref{fig: result}A), it yielded lower perceptual similarity ratings for some textures (Figure~\ref{fig: Result_tex}), likely because the other representations had greater spectral energy after normalization while using the same rendering gain as AR. Previous work has shown that increased low- and high-frequency energy enhances the perception of bumpiness and roughness~\cite{hollins2007coding, bau2010teslatouch}, which may explain the higher similarity ratings observed for most methods on coarse textures. sPeak performed slightly worse for coarse fabric textures, likely because its limited number of spectral peaks could not fully capture both low- and high-frequency components of the signal. sSlope produced lower perceptual similarity ratings than sPeak for some textures (Figure~\ref{fig: Result_tex}), possibly due to its low- and high-pass filter orders which were quantized to the nearest 20~dB/decade, thereby amplifying intermediate-frequency components between the spectral peak and the tactile frequency limits. In contrast, fine textures and vinyl were reproduced effectively by most methods, except MFCC (Figure~\ref{fig: Result_tex}), as their dominant high-frequency components (300–600~Hz) were captured reliably and further attenuated naturally during electrovibration interaction~\cite{balasubramanian2025sliding}. MFCC consistently produced the lowest perceptual similarity ratings (Figure~\ref{fig: result}B), likely because its Mel-scale filterbank, originally designed for auditory perception, does not adequately represent tactile frequency content.

% Linear regression
Previous studies have shown that frictional and vibratory stimuli are processed in multiple frequency-tuned channels that integrate spectral energy to form tactile perception~\cite{makous1995critical, bensmaia2005vibrotactile, kuroki2017integration}. Consistent with this framework, our linear regression analysis suggests that the proposed representations effectively capture texture information distributed across nine critical frequency bands for the tested texture set (Figure~\ref{fig: result}C). Although the regression metrics (p-value, $r^2$, and RMSE) reached a plateau at nine filters, the absolute $r^2$ values remain relatively low. One possible explanation is that the texture recordings were obtained from the first author rather than from the participants themselves, thereby introducing subject-specific differences in skin mechanics. Furthermore, variability in measured contact forces arising from finger-electrovibration interactions~\cite{balasubramanian2025sliding, kenanoglu2026adhesion} and fluctuations in skin hydration ~\cite{aliabbasi2024effect} likely contributed additional variance. Together, these findings suggest that further investigation under more tightly controlled experimental conditions is needed to more precisely identify the critical frequency bands underlying effective tactile texture representation.

% Limitation and future work
%Despite taking utmost care during experiments, we still encountered certain limitations. First, texture recordings obtained from the first author were used for rendering, creating a mismatch because the recorded friction does not reflect each participant’s individual contact forces, lowering the similarity. Second, experiments were conducted at a single exploration speed and normal force, limiting generalizability. Furthermore, the use of active exploration introduced variability in individual finger exploration, potentially influencing measured contact forces and subsequent analyses. 

%conclusion statement.
% In the future, we will address current limitations by extending the evaluation to a broader range of textures and exploration conditions. Additionally, we aim to improve rendering fidelity by modeling texture spectra within a reduced set of perceptually relevant subbands, as indicated by our linear regression analysis, and validate this approach through psychophysical studies.

% In conclusion, XXX...

%In conclusion, the proposed sBeta representation achieved perceptual similarity comparable to high-fidelity texture recordings while requiring only two parameters. 

In conclusion, our findings introduce compact texture representations that can potentially enable efficient texture storage, communication, and the synthesis of new textures without extensive real-world recordings. More broadly, the results suggest that humans perceive fine textures primarily through fundamental temporal spectral patterns across multiple frequency bands, and that modeling these patterns is
sufficient to reproduce perceptually realistic tactile
sensations.

%Future work can leverage these parameters for AI-driven texture classification and perception in robotic touch applications.

\section{Materials and Methods}

% \begin{figure}[!ht]
%     \centering
%     \includegraphics[width=\linewidth]{Data processing.pdf}
%     \caption{(A) Schematic of the setup for recording lateral friction force ($F_{L}$) generated during finger–surface scanning at a normal force ($F_N$) = 0.4~N and speed ($\nu$) = 80~mm/s; the texture mounted on an acrylic base was attached to a force sensor. (B) Recorded lateral force signal (yellow), extracted during lateral sweeps (red), filtered to the human exploration bandwidth (10–1000 Hz), compensated for the setup's response ($H_{rec}^{-1}(s)$), and transformed to the frequency domain. (C) Texture representations: autoregression (AR), Mel-frequency cepstral coefficients (MFCC), spectral peak (sPeak), and the proposed spectral beta (SBeta) and spectral slope (sSlope); model outputs were compensated for the finger electrovibration response ($H_{EV}^{-1}(s)$) and amplitude-modulated with a high-frequency carrier after square root operation. (D) Electrovibration rendering setup, where the processed model outputs excite the touchscreen to generate lateral friction ($\hat{F_{L}}$).}
%     \label{fig: dataprocess}
% \end{figure}

\subsection*{Texture recording and processing}
The texture representations were retrieved from recording finger-surface friction signals from five natural textures: fine fabric, coarse fabric, corrugated paper, sandpaper, and vinyl, which are included in the SENS3 texture database~\cite{balasubramanian2024sens3} (Fig.~\ref{fig: modelOut})E. Each texture sample (100$\times$100~mm) was affixed to an acrylic plate of the same size and mounted on a six-axis force sensor (Nano 43, ATI Inc.), which recorded interaction forces at 20~kHz via a data acquisition card (PCIe-6321, NI Inc.). The first author scanned each texture laterally using the dominant index finger at a commonly preferred scanning speed of 80~mm/s, applying a normal force of 0.4~N~\cite{kejriwal2023user} for 10 seconds at a 60° contact angle (Fig.~\ref{fig: modelOut}A). An infrared position sensor (NNAMC1580PCEV, Neonode) sampling at 60~Hz tracked finger position and speed.

After recording, the lateral-sweep friction data were extracted and transformed to the frequency domain as shown in Fig.~\ref{fig: modelOut}B. The x- and y-axis force sensor measurements were combined into a 1D friction signal~\cite{landin2010dimensional}. Lateral sweep data segments were extracted using mean thresholding, windowed with a Hanning function for seamless stitching, and averaged over the 10-second recording, with 4000-sample segments providing 5~Hz frequency resolution. The signals were bandpass filtered between 20~Hz and 1~kHz to remove low-frequency motion artifacts and high-frequency components attenuated by the finger~\cite{morioka2005thresholds}, followed by compensation for the recording setup’s response; see Supplementary Information for details.

\subsection*{Texture models}
We proposed two new texture representations, spectral beta and spectral slope, and compared them with state-of-the-art models (Auto regression, Mel Frequency Cepstral Coefficients, and Spectral Peak (Fig.~\ref{fig: modelOut}C) 

\subsubsection*{Spectral beta}
The spectral beta (sBeta) models the envelope of a texture’s magnitude spectrum using a beta distribution~\cite{hahn1967statistical}. Visually, texture spectra often resemble statistical distributions, motivating the choice of the beta distribution, which is parameterized by only two shape parameters: $\alpha$ and $\beta$. Its ability to represent both left- and right-skewed distributions makes it well suited for modeling textures dominated by low-, mid-, or high-frequency content. Since the Beta distribution is defined over the interval $[0,1]$, the log frequency axis of the input spectrum was normalized before curve fitting. The ten spectral peaks were used as reference data points to fit the Beta distribution using MATLAB's fminsearch algorithm to estimate the parameters $\alpha$ and $\beta$. Using a larger number of spectral points may lead to overfitting. The fitted model was then mapped back to the original frequency scale and multiplied by a rectangular window spanning 20–1000~Hz to confine energy to the tactile range (Fig.~\ref{fig: modelOut}C). The resulting magnitude spectrum was combined with a random phase and transformed to the time domain to synthesize the texture signal, as phase differences are generally not perceived in touch for homogeneous textures~\cite{kuroki2021roles}.

\subsubsection*{Spectral slope}
The spectral slope (sSlope) representation models the texture spectra using an asymmetric triangular bandpass filter centered at the dominant spectral peak. The bandpass comprises a high-pass and a low-pass filter, both with the same peak frequency as the cutoff. Slopes were estimated from 20~Hz to the peak and from the peak to 1~KHz. These rates were rounded to the nearest 20~dB/decade and used to determine the filter orders of the high-pass ($r_a$) and low-pass ($r_b$) components (Fig.~\ref{fig: modelOut}C). Using the orders as texture representations, an asymmetric narrow bandpass filter was constructed and applied to white noise to synthesize the texture signals.

\subsubsection*{Auto regression, Mels Frequency Cepstral Coefficient, and Spectral Peak}
We compared the sBeta and sSlope representations with state-of-the-art texture models, including Auto Regression (AR), Mel Frequency Cepstral Coefficients (MFCC), and Spectral Peaks (sPeak). The AR method models texture signals as time series using an all-pole infinite-impulse-response filter, with the resulting coefficients serving as the representation~\cite{culbertson2014modeling}. MFCC represents spectral content using a perceptually spaced Mel filterbank, specified for audio perception~\cite{vergin2002generalized}, and encodes the power in each band. It has been applied to tactile feature extraction and texture synthesis~\cite{ strese2014haptic, hassan2020authoring}. The encoded band powers serve as the representation and are used to reconstruct the signal. The sPeak method extracts spectral peaks separated by approximately 12\% just-noticeable difference (JND), typically selecting $~$10 peaks; each peak’s frequency and magnitude generate sinusoids whose sum reconstructs a perceptually lossless texture signal~\cite{fielder2019novel}. Additional details on the representations are provided in the Supplementary Information.

\subsection*{Texture rendering}

Reconstructed signals from the models were first preprocessed for electrovibration rendering by compensating for high-frequency attenuation caused by the mechanical interactions between the finger and the electrostatic screen~\cite{balasubramanian2025sliding} (Fig.~\ref{fig: modelOut}C). The signals were then normalized, square-root transformed, and amplitude-modulated with a 7~kHz carrier, following established rendering procedures~\cite{balasubramanian2025sliding}. 

\subsection*{Experiment}
The experiment was conducted with 14 participants (9 men and 5 women; mean age: 28 years, SD = $\pm$3.34). The sample size was determined through power analysis (see Supplementary Information for more details), and all procedures were performed in accordance with the ethical principles of the Declaration of Helsinki. Ethical approval was obtained from the TU Delft Human Research Ethics Committee (application number 3469). All participants provided informed consent before participation.

\begin{figure}[!ht]
    \centering
    \includegraphics[width=\linewidth]{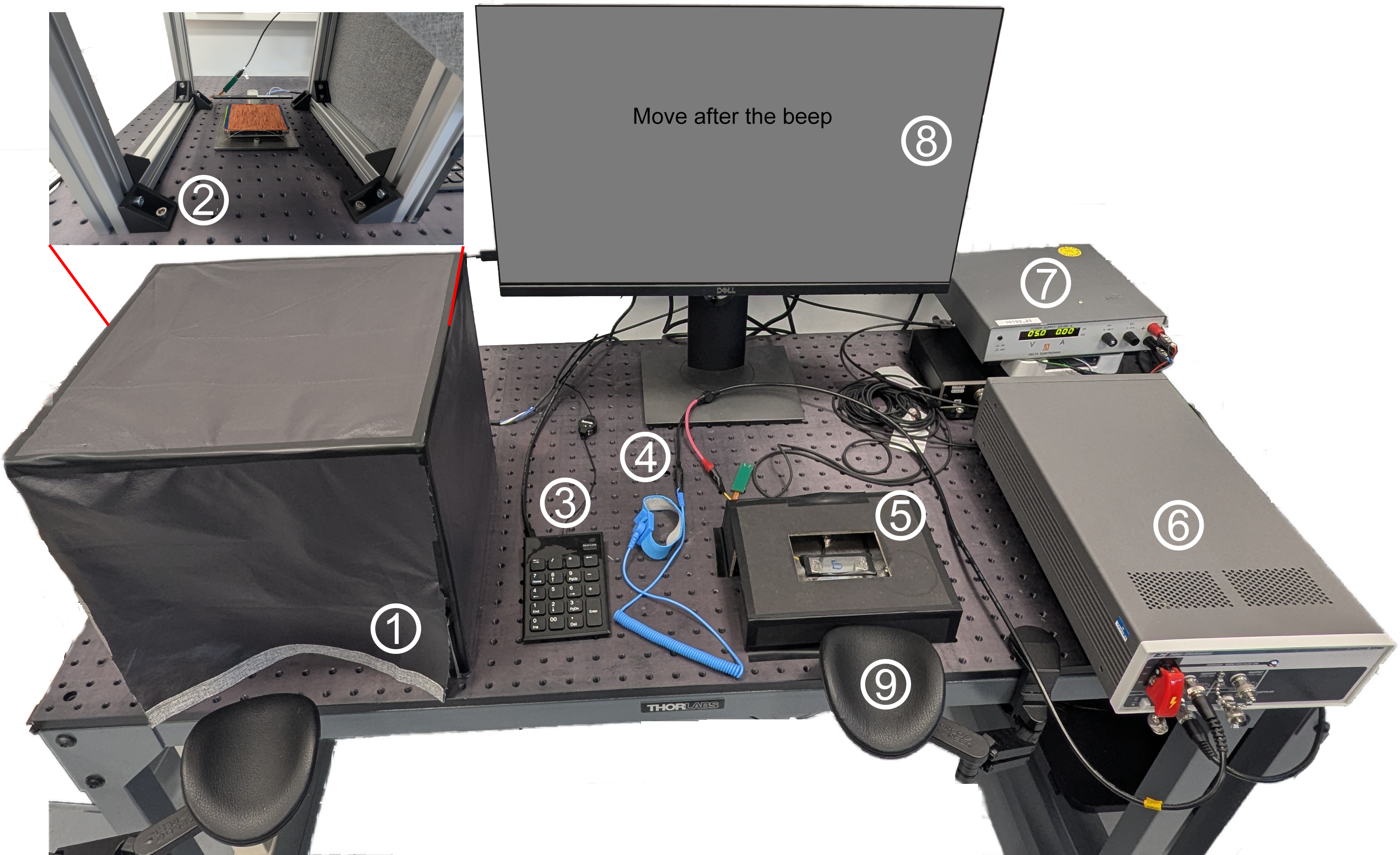}
    \caption{\textbf{Illustration of experimental setup} (1) Real texture cover, (2) Internal view showing the real texture mounted on a load cell, (3) Keypad for recording similarity ratings, (4) Grounding strap, (5) 3M touchscreen mounted on force sensors, (6) High-voltage amplifier, (7) Power supply for the load cell, (8) Monitor for visual feedback, (9) Armrest for participant support.}
    \label{fig: Setup}
\end{figure}

Participants simultaneously explored a virtual texture on a capacitive touchscreen (SCT3250, 3M Inc.) with the right index finger and the corresponding real texture with the left index finger, both at a $60^\circ$ contact angle (Fig.~\ref{fig: concept}). The touchscreen was mounted on two six-axis force sensors (Nano17, ATI Inc.) and excited with signals generated via a data-acquisition card (PCIe-6321, NI Inc.) after high-voltage amplification (9200A, Tabor Inc.). Sliding speed was measured at 60~Hz using an infrared position sensor (NNAMC1580PCEV, Neonode). The visually occluded real texture was mounted on a load cell (YZC-131, 1 kg). All measured force signals were sampled at 20~kHz. Schematics of the rendering and the full experimental setup are shown in Figure~\ref{fig: modelOut}D and~\ref{fig: Setup}.

Before the experiment, participants washed and dried their hands. Then, they wore a grounding strap on the right hand. The experiment consisted of three consecutive phases: training, gain-matching, and testing. Throughout all these phases, the participants wore noise-canceling headphones and listened to white noise to mask external auditory cues and those caused by finger-surface interactions. In the training phase, participants practiced maintaining a scanning speed of 80~mm/s and an applied normal force of 0.4~N while sliding over both the real texture and the touchscreen in the medial-to-lateral direction. An auditory cue prompted participants to synchronize their scanning speed to a metronome and regulate force using a graphical user interface with color-coded feedback (red: above target, yellow: below target, green: on target). In the gain-matching phase, participants adjusted the touchscreen input gain such that the virtual texture modeled by Auto regression (AR) was perceived as similar to the real one in terms of friction and roughness intensity~\cite{grigorii2021data}. Gain adjustment began at the maximum voltage level and was reduced until perceptual equivalence was reported. AR representation was used for calibration because it exhibited the highest spectral correlation with the original texture. The calibrated gain was then applied to all other texture representations. During testing, participants scanned both the real and the corresponding virtual textures and rated their similarity on a 7-point Likert scale (0 = “Different,” 6 = “Same”). Five representations were evaluated across five textures, with five repetitions per combination. Trials and sessions lasted 10 seconds and $~$2.5 hours per participant, respectively.

During the experiments, contact forces from interactions with the virtual textures were recorded for analysis. As the setup included two force sensors, the x, y, and z components of each sensor were first summed to obtain a single 3D force vector. Following the procedure used for the recorded real textures, the lateral components (x and y axes) were combined to generate a one-dimensional friction signal. Segments of 4000 samples were then extracted using a Hanning window for analysis.

\section*{Data availability}
The data and code used in this paper will be publicly available upon acceptance in the \href{http://dx.doi.org/10.4121/e7858a25-752e-4afd-82cd-895297c10bcc}{4TU.ResearchData} repository.

\section*{Funding statement}
This work was partly funded by the Dutch Research Council (NWO) with project number 19153.

\section*{Supplementary}

\begin{figure}[!ht]
    \centering
    \includegraphics[width=\linewidth]{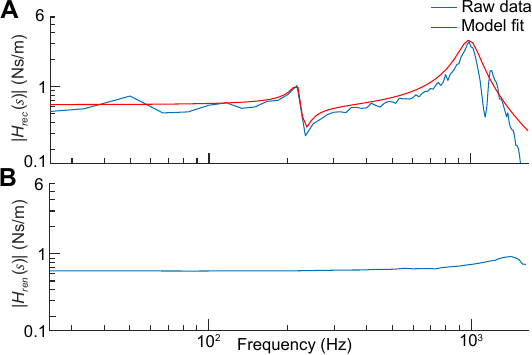}
    \caption{\textbf{Frequency responses.}(A) Frequency response of the recording setup in the lateral direction. (B) Frequency response of the rendering setup in the lateral direction. Both setups were excited using an impact hammer, and the normal-direction response was measured with a force sensor.}
    \label{fig: reponse}
\end{figure}

\subsection{Recording setup}
We built a custom data-collection setup similar to the one used in SENS3~\cite{balasubramanian2024sens3} to record textures via contact forces generated during the first author's interactions with the textures. The setup, consisting of a force sensor (Nano17 Titanium, ATI) mounted on an aluminum block, was placed on an aluminum optical breadboard table (MB60120/M, Thorlabs) and mounted onto a robust supporting frame (PFM52502, Thorlabs). A 3~mm-thick acrylic base plate was placed above the sensor, onto which 100×100~mm texture samples were attached. A data-acquisition card (NI PCIe-6323) collected data from the force sensor at a sampling rate of 20~kHz. We used a 2D infrared position sensor (NNAMC1580PCEV, Neonode) with a 0.1mm resolution and a 60 Hz sampling rate to measure the index finger speed. The sensor was attached using 3D-printed support, and the active plane of the sensor was positioned above the material. The entire setup was covered with a cardboard box with a cutout exposing the textures. A monitor (Dell UltraSharp 24) displayed graphical information to the participants. Texture recording and processing were performed on a PC (Precision T5820, Dell). 

\subsection{Texture recording procedure}\label{sec: texREc}
We aimed to capture the contact forces generated during finger-surface interaction. To ensure consistency, the first author collected all the texture recordings by exploring the textures using their dominant hand index finger with 60$^{\circ}$ contact angle. The cardboard enclosure with a finger exploration cutout maintained the contact angle throughout the exploration. For each texture, the author performed left-to-right (lateral) and vice versa (medial) finger sweeps at a scanning speed of 80~mm/s with a force of 0.4~N for 10 seconds. A custom-built graphical user interface (GUI) guided the author in meeting the exploration conditions. The interface featured two squares: the top square displayed the target scanning speed, while the bottom square tracked the author’s finger motion in real-time. The tracking square also changed color based on the applied force—yellow indicated the force was below the target range, green indicated it was within the target range, and red indicated it exceeded the target range. If the mean, minimum, or maximum values of the scanning speed or applied force deviated by more than $\pm10\%$ from the target, the recording was discarded and repeated.

\subsection{Texture processing}\label{sec: texPro}
After the texture recording, we extracted and processed the desired friction signal and filtered it for hand motion before modeling. Friction from finger movement was mainly captured along the lateral direction by the y-axis of the force sensor, though slight deviations in finger trajectory introduced components along the x-axis. We combined the data from both axes by reducing the 2D signal to a 1D lateral force~\cite{landin2010dimensional}. We computed the mean of the resultant friction signal and used it as a threshold to detect finger sweeps. The friction signal above this threshold was considered a lateral sweep, and we found its start and end indices. To minimize spectral leakage during data extraction, we applied a 4000-sample-long Hanning window, centered at the midpoint between the start and end indices of each sweep, yielding 4000 samples per sweep. The 4000 samples provided a 5~Hz frequency resolution in the FFT, which is necessary for accurately modeling the texture spectrum. This process was repeated for all detected sweeps. We computed the FFT of each sweep’s friction signal and averaged the FFTs to reduce noise and improve the signal quality of the recorded texture signals. Finally, we applied a bandpass filter of 20~Hz to 1~kHz to the texture signals to eliminate low-frequency trends from hand motion, as well as high-frequency components above 1~kHz, which are naturally attenuated by the finger~\cite{morioka2005thresholds}.

We then removed the setup’s natural response (shown in Fig.~\ref{fig: reponse}A) from each texture signal. To find the setup's natural response, we struck the base plate with texture along its lateral axis using an impact hammer (086E80, PCB Piezotronics). We divided the FFT of the force sensor's y-axis response by the FFT of the impact hammer response to get the setup's response. Since the force sensor in the setup had equal stiffness along both axes, the x-axis response was found to be similar to the y-axis. Finally, we corrected each texture signal by dividing its FFT by the setup’s response. The supplementary section in~\cite{balasubramanian2025sliding} for more details on the setup's response compensation.

\begin{figure}[!ht]
    \centering
    \includegraphics[width=\linewidth]{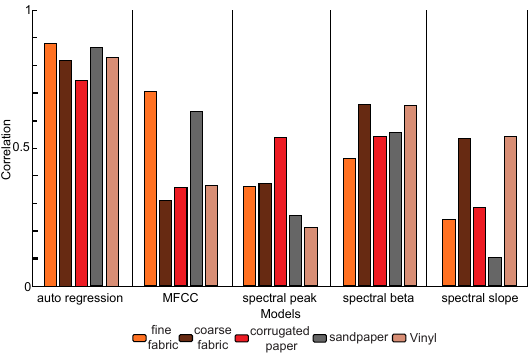}
    \caption{Illustration showing the correlation between original texture recordings and representation outputs prior to rendering.}
    \label{fig: modelCorr}
\end{figure}

\subsection{Texture modeling and rendering}
This section explains autoregression, mel-frequency cepstral coefficients, and spectral peak representations.

\subsubsection{Auto-regression}
Autoregressive (AR) modeling represents friction signals as time series using an all-pole infinite-impulse-response (IIR) filter, where the current output is expressed as a linear combination of past outputs. The structure of an AR model is given as:

\begin{equation}
    y_{ar}(t) = \sum_{i=1}^{p}A(i)y_{ar}(t-i) + \epsilon(t),
\end{equation}

Where $y_{ar}(t)$ and $y_{ar}(t-i)$ are current and past outputs, $p$ is the model order, $A(i)$ are the AR coefficients, and $\epsilon$ is the residual value. Following the approach of Culbertson et al.~\cite{culbertson2014modeling}, the six AR coefficients were estimated from the recorded friction signals, and the model order was kept uniform for all textures.

The estimated order, $p$, and AR coefficients, $A(i)$, were then used to construct a discrete-time filter to synthesize texture signals by filtering white noise. The corresponding transfer function, $H_{ar}(z)$, is given by:

\begin{equation}
    H(z) = \frac{1}{1-\sum_{k=1}^{p}A(k)z^{-k}}
\end{equation}

\subsubsection{Mel Frequency Cepstral Coefficients}
Mel-frequency cepstral coefficients (MFCCs) are widely used for audio feature extraction~\cite{vergin2002generalized} and have been applied to tactile signal analysis~\cite{strese2014haptic}. In this approach, friction signals are transformed to the frequency domain, mapped onto the perceptually motivated Mel scale, and passed through a set of overlapping Mel filter banks. The filter-bank energies after a discrete cosine transform (DCT) yield the MFCC features.

Although MFCCs primarily encode spectral magnitude information and do not preserve phase, approximate signal reconstruction is possible. Based on prior work indicating that 8–14 coefficients are sufficient to represent audio signals~\cite{abdouni2018impact}, we retained 10 coefficients. Reconstruction was performed by applying the inverse DCT, followed by inverse Mel filtering, inverse Fourier transform, and conversion from the Mel to linear frequency scale. Because MFCCs encode only magnitude information, a random phase was assumed during reconstruction.

\subsubsection{Spectral peak}
For the spectral peak method, we followed the procedure proposed by Fielder and Vardar~\cite{fielder2019novel}, which represents texture signals using dominant frequency components extracted from the magnitude spectrum. After computing the fast Fourier transform (FFT) of the filtered texture signals, prominent spectral peaks were selected using a 12\% just-noticeable difference (JND) criterion, enabling perceptually lossless compression by retaining only distinguishable frequency components. The reconstructed signal was synthesized as a sum of sinusoids corresponding to the selected peaks:
\begin{equation}
    y_{sp}(t) = \sum_{i = 1}^{n} A_{i} \cos (2\pi f_{i}t)
    \label{eqn: Spectral}
\end{equation}
where, $y_{sp}(t)$ is the output of the spectral peak model, $A_i$ is the magnitude of the $i$th peak component, $f_i$ is the frequency, and $n$ is the total number of peaks. Consistent with prior findings that up to 10 frequency components are sufficient to approximate most real-world textures~\cite{fielder2019novel}, we set 
$n$ = 10 in this study.

\subsection{Electrovibration rendering}
The outputs of each representation are rendered on an electrovibration display, similar to the technique shown in~\cite{balasubramanian2025sliding}. The model outputs are DC-shifted and amplitude-modulated to avoid problems related to electrovibration, such as nonlinearity and charge leakage. The input voltage signal for electrovibration, $V(t)$, based on the model output, $y(t)$ is given as:

\begin{equation}
    V(t) = V_g\sqrt{y(t)+min|y(t)|}cos(2 \pi f_c t)
\end{equation}
here, $f_c$ is the carrier frequency of 7~kHz for amplitude modulation. 

Finally, the participants adjusted the output signal gain, $V_g$, to match the roughness sensations of the rendered textures with those of their corresponding real-world counterparts.

\subsection{Gain matching task}
Participants adjusted the gain, $V_g$, of the rendered textures to ensure that the friction-induced vibrations matched those of the corresponding real textures~\cite{grigorii2021data}. The textures generated using the auto-regressive model served as the reference textures for rendering. The gain obtained for the reference texture was applied to textures generated using the other representations.

We used the methods of adjustment to find the gain, where the first author controlled the gain of the rendered signal using a slider. The gain started with a maximum value which corresponded to a peak output voltage of 150V. During the experiment, participants compared the textures rendered on the touchscreen with their real-world counterparts by scanning the touchscreen using the dominant index finger while maintaining an applied force of 0.4 N and a scanning speed of 80 mm/s~\cite{kejriwal2023user}. Participants then reported whether the rendered texture produced a higher or lower perceived friction and roughness sensation relative to the real texture. Based on this feedback, the first author adjusted the slider to increase or decrease the gain. The trial ended when participants reported that the rendered and real textures felt similar. This procedure was repeated for all five textures: fine fabric, coarse fabric, corrugated paper, sandpaper, and vinyl.

\subsection{Power analysis}
A prior power analysis was conducted following the guidelines of Cohen~\cite{cohen2016power} using GPower 3.1 software to determine the required sample size. We selected the F-test family with a repeated-measures ANOVA (within-factors design). The effect size was set to 0.25 based on preliminary results from the first five participants. The significance level, $\alpha$, was set to 0.05 and the statistical power (1-4$\alpha$) to 0.8. With one group and 25 measurements (5 textures × 5 models), the analysis indicated a sample size of 8 participants was required. To account for the use of nonparametric statistical tests and to improve the reliability of the results, the sample size was increased to 14 participants.

\subsection{Linear regression}
To investigate how the representations contributed to the perceived similarity between the rendered and real textures, we performed a linear regression analysis. The model used the band-wise spectral energy differences between the measured finger–touchscreen friction signals and the original texture signal energy as independent variables. Participants' subjective similarity ratings were used as the dependent variable. Prior to linear regression, we averaged the measured friction signals in the frequency domain and the subjective ratings across repetitions for each participant and texture-representation combination to reduce measurement noise. We then calculated the frequency-wise absolute spectral energy difference between the measured contact forces of the original and rendered textures. Smaller energy differences indicated that the rendered textures more closely captured the original texture characteristics.

Next, we standardized the energy differences and average subjective ratings for each texture within each participant using z-scores to account for individual differences in finger mechanics and response biases. The standardized energy differences within the tactile frequency range (20–1000 Hz) were divided using triangular filter banks on a logarithmic frequency scale. The first filter bank consisted of a single triangular filter spanning the entire tactile frequency range, and each subsequent filter bank added one filter. Within each filter bank, the filters are equally subdivided to cover the tactile frequency range with 50\% overlap. This process was repeated up to 20 times.

The standardized average subjective ratings served as the dependent variable, while the total energy difference within each filter was used as the input feature for the linear regression. Across the 20 iterations, the regression metrics—including p-value, $r^2$, F-statistic, and root mean square error (RMSE)—plateaued at nine filters.

\bibliographystyle{unsrt}
\bibliography{Reference}  

\end{document}